\documentclass[
    ,final            
  ]
  {aipproc}

\usepackage{epsfig}
\usepackage{latexsym}
\usepackage{graphicx}
\usepackage{amsmath}
\usepackage{amsfonts}   
\usepackage{amssymb}    
\layoutstyle{6x9}

\newcommand{\newc}{\newcommand}
\newc\eg{{\it {e.g.}}}  \newc\etal{{\it {et al.}}} \newc\ie{{\it i.e.}}
\newc\etc{{\it {etc}}}  

\newcommand\lsim{\mathrel{\rlap{\lower4pt\hbox{\hskip1pt$\sim$}}
    \raise1pt\hbox{$<$}}}
\newcommand\gsim{\mathrel{\rlap{\lower4pt\hbox{\hskip1pt$\sim$}}
    \raise1pt\hbox{$>$}}}

\newc{\sigsip}{\sigma^{SI}_{p}}	\newc{\sigsin}{\sigma^{SI}_{n}}
\newc{\sigsdp}{\sigma^{SD}_{p}}	\newc{\sigsdn}{\sigma^{SD}_{n}}
\newc{\sigsi}{\sigma^{SI}}	\newc{\sigsd}{\sigma^{SD}}

\newc{\mhalf}{m_{1/2}}      \newc{\mzero}{m_0}
\newc{\tanb}{\tan\beta}
\newc{\azero}{A_0}
\newc{\at}{A_t} \newc{\abot}{A_b} \newc{\atau}{A_\tau} 

\newc{\bmu}{B\mu}           \newc{\sgn}{{\rm sgn}}
\newc{\mone}{M_1}           \newc{\mtwo}{M_2}

\newc{\charone}{\chi_1^\pm} \newc{\mcharone}{m_{\chi_1^\pm}}

\newc{\hl}{h}               \newc{\mhl}{m_{\hl}}
\newc{\hh}{H}               \newc{\mhh}{m_{\hh}}
\newc{\ha}{A}               \newc{\mha}{m_{\ha}}
\newc{\hc}{H^{\pm}}         \newc{\mhc}{m_{\hc}}

\newc{\mw}{m_{W}}      \newc{\mz}{m_{Z}}

\newc{\mgut}{M_{\rm GUT}}

\newc{\mplanck}{M_{\rm P}}      \newc{\mpl}{M_{\rm Pl}}
\newc{\msusy}{M_{\rm SUSY}}      \newc{\ms}{M_{\rm S}}

\newc{\jxf}{J({\xf})}
\newc{\jxfexact}{J_{\rm exact}({\xf})}  \newc{\jxfexp}{J_{\rm exp}({\xf})}
\newc{\VEV}[1]{\langle #1 \rangle}

\newc{\xf}{x_f}
\newc\vrel{v_{\rm rel}}
\newcommand\mchi{m_{\chi}}              

\newc\sell{{\widetilde e}_L}      \newc\msell{m_{\sell}}
\newc\selr{{\widetilde e}_R}      \newc\mselr{m_{\selr}}

\newc\snue{{\widetilde \nu}_e}      \newc\msnue{m_{\snue}}
\newc\snutau{{\widetilde \nu}_\tau}      \newc\msnutau{m_{\snutau}}

\newc\supl{{\widetilde u}_L}      \newc\msupl{m_{\supl}}
\newc\supr{{\widetilde u}_R}      \newc\msupr{m_{\supr}}
\newc\sdl{{\widetilde d}_L}      \newc\msdl{m_{\sdl}}
\newc\sdr{{\widetilde d}_R}      \newc\msdr{m_{\sdr}}

\newcommand{\stau}{{\tilde \tau}}   \newcommand\mstau{m_{\stau}}

\newcommand\gluino{\tilde g}
\newcommand\mgluino{m_{\gluino}}

\newc\hpm{H^\pm} \newc\hp{H^+} \newc\hm{H^-} 
\newc\sfermion{\tilde f}  \newc\msfermion{m_{\sfermion}}  

\newc\alphas{\alpha_s}

\newc\alphaem{\alpha_{em}}

\newcommand\treh{T_{\rm R}}     \newcommand\trehmax{T_{\rm R}^{\rm max}}

\newc{\sthw}{\sin\theta_W}       \newc{\ssqthw}{\sin^2\theta_W}         
\newc{\cthw}{\cos\theta_W}       \newc{\csqthw}{\cos^2\theta_W}
\newc{\tthw}{\tan\theta_W}       \newc{\tsqthw}{\tan^2\theta_W}       

\newc{\bino}{\widetilde B}              \newc{\wino}{\widetilde W_3}
\newc{\higgsinob}{{\widetilde H}^0_b}   \newc{\higgsinot}{{\widetilde H}^0_t}

\newc{\abund}{\Omega h^2}               \newc{\abundobs}{\Omega_{\rm obs} h^2}

\newc{\abundchi}{\Omega_\chi h^2}
\newc{\abundcdm}{\Omega_{{\rm CDM}} h^2}

\newc{\omegam}{\Omega_{{\rm M}}}       \newc{\abundm}{\Omega_{{\rm M}} h^2}
\newc{\omegab}{\Omega_{{\rm b}}}	\newc{\abundb}{\Omega_{{\rm b}} h^2}
\newc{\omegatot}{\Omega_{{\rm TOT}}}

\newc{\nlsp}{n_{{\rm LSP}}} \newc{\mlsp}{m_{{\rm LSP}}} \newc{\mlspmax}{m_{\rm LSP}^{\rm max}}     
\newc{\ylsp}{Y_{{\rm LSP}}} 
\newc{\abundlsp}{\Omega_{\rm LSP}h^2}
\newc{\abundlsptp}{\Omega_{\rm LSP}^{\rm TP}h^2} 
\newc{\abundlspntp}{\Omega_{\rm LSP}^{\rm NTP}h^2}

\newc{\omeganlsp}{\Omega_{{\rm NLSP}}}   
\newc{\ynlsp}{Y_{{\rm NLSP}}}            \newc{\taunlsp}{\tau_{{\rm NLSP}}}
\newc{\nnlsp}{n_{{\rm NLSP}}}            \newc{\mnlsp}{m_{{\rm NLSP}}}
\newc{\abundnlsp}{\Omega_{\rm NLSP}h^2}
\newc{\abundnlsptp}{\Omega_{\rm NLSP}^{\rm TP}h^2} 
\newc{\abundnlspntp}{\Omega_{\rm NLSP}^{\rm NTP}h^2}

\newc{\nx}{n_{X}}                        \newc{\yx}{Y_{X}}
\newc{\mx}{m_{X}}                        \newc{\taux}{\tau_{X}}

\newc{\rhocrit}{\rho_{crit}}
\newc{\rhochi}{\rho_{\chi}}

\newcommand\fa{f_{a}}

\newcommand\neut{\tilde \chi}

\newc{\cachigamma}{C_{a\neut\gamma}}
\newc{\caww}{C_{aWW}}                   
\newc{\cayy}{C_{aYY}}

\newc{\nl}{\cos \theta_{\tilde t}}
\newc{\nr}{\sin \theta_{\tilde t}}
\newcommand\tev{\,{\rm TeV}}
\newcommand\gev{\,{\rm GeV}}

\newcommand\kev{\,{\rm keV}}

\newc\gbar{{\overline{g}}}

\newc{\ra}{\rightarrow}

\newc{\beq}{\begin{equation}}
\newc{\eeq}{\end{equation}}
\newc{\bea}{\begin{eqnarray}}
\newc{\eea}{\end{eqnarray}}
\newcommand{\beqa}[1]{\begin{eqnarray}#1\end{eqnarray}}

\newc{\nspin}{n_{\rm spin}}
\newc{\nflavor}{n_{\rm F}}
\newc{\ngamma}{n_\gamma}
\newc{\ychi}{Y_{\chi}}                  \newc{\yeqchi}{Y^{\rm EQ}_{\chi}}

\newcommand\axino{{\tilde{a}}}        
\newcommand\maxino{{m_{\axino}}}

\newcommand\abunda{\Omega_{\axino}h^2}

\newcommand\abundatp{\Omega^{\rm TP}_{\axino}h^2}       

\newc{\naxino}{n_{\axino}}
\newc{\yaxino}{Y_{\axino}}
\newc{\yaxinoeq}{Y^{\rm EQ}_{\axino}}
\newc{\yaxinotp}{Y^{\rm TP}_{\axino}}
\newc{\yaxinontp}{Y^{\rm NTP}_{\axino}}

\newcommand\gravitino{{\widetilde{G}}}    
\newcommand\mgravitino{{m_{\gravitino}}}

\newcommand\abundg{\Omega_{\gravitino}h^2}

\newcommand\abundgtp{\Omega^{\rm TP}_{\gravitino}h^2}       

\newc{\ngravitino}{n_{\gravitino}}
\newc{\ygravitino}{Y_{\gravitino}}
\newc{\yeqgravitino}{Y^{\rm EQ}_{\gravitino}}
\newc{\ygravitinotp}{Y^{\rm TP}_{\gravitino}}
\newc{\ygravitinontp}{Y^{\rm NTP}_{\gravitino}}

\newc{\yascat}{Y^{\rm scat}_{i,j}}      \newc{\yadec}{Y^{\rm dec}_{i}}
\newc{\gstar}{g_\ast}           \newc{\gsstar}{g_{s\ast}}

       \def\pslash{\not{\hbox{\kern-2.3pt $p$}}}
       \def\kslash{\not{\hbox{\kern-2.3pt $k$}}}
       \def\qslash{\not{\hbox{\kern-2.3pt $q$}}}
       \def\ddslash{\not{\hbox{\kern-2.3pt $d$}}}
       \def\prtslash{\not{\hbox{\kern-2.3pt $\partial$}}}

\begin{document}

\title{ 	
 Determining Reheating Temperature at LHC with Axino or Gravitino Dark Matter\footnote{Talk given by K.-Y. Choi at International Workshop on Dark Side Of the Universe, 1-5 Jun 2008, Cairo, Egypt.}}


\author{Ki-Young Choi}{
  address={
        Departamento de F\'{\i}sica Te\'{o}rica C-XI
        and Instituto de F\'{\i}sica Te\'{o}rica UAM/CSIC,\\
        Universidad Aut\'{o}noma de Madrid, Cantoblanco,
        28049 Madrid, Spain\\
}}
\author{Leszek Roszkowski}{
  address={
        Department of Physics and Astronomy, University of Sheffield,\\
        Sheffield S3 7RH, England\\
}}
\author{Roberto Ruiz de Austri}{
  address={
        Departamento de F\'{\i}sica Te\'{o}rica C-XI
        and Instituto de F\'{\i}sica Te\'{o}rica UAM/CSIC,\\
        Universidad Aut\'{o}noma de Madrid, Cantoblanco,
        28049 Madrid, Spain\\
}}

\begin{abstract}
 After a period of inflationary expansion, the
  Universe reheated and reached full thermal equilibrium at the
  reheating temperature $\treh$. In this talk, based on~\cite{Choi:2007rh}, 
  we point out that, in
  the context of effective low-energy supersymmetric models, LHC
  measurements may allow one to determine $\treh$ as a function of the
  mass of the dark matter particle assumed to be either an axino or a
  gravitino. An upper bound on their mass and on $\treh$ may also be
  derived.
\end{abstract}

\maketitle


\section{Introduction}
Recent astrophysical and cosmological observation
 give precise determination on the relic
density of cold dark matter in the range~\cite{Spergel:2006hy}
$\abundcdm =0.104 \pm 0.009 $.
The well motivated candidate for dark matter (DM) is weakly interacting
massive particle (WIMP), especially the lightest neutralino.
In addition to the direct and indirect experiments to explore WIMP,
the Large Hadron Collider has already started to 
reveal the secret of TeV energy
and is expected to find several superpartners and to determine their
properties. In particular, the feasibility was investigated to determine 
the neutralino's mass $m_\chi$~\cite{mass,Barr:2008ba} and relic abundance 
$\abundchi$~\cite{oh2atlhc-cmssm,oh2atlhc-mssm} from LHC
measurements. 
An analogous study has also been done in the
context of the Linear Collider, where accuracy of a similar
determination would be much better~\cite{oh2atilc}.

However neutralino is found as
an apparently stable state in LHC detectors, but may not be the true LSP
and therefore not DM in the Universe. Instead, it could decay 
into an even lighter and weakly interacting state, the real LSP, outside the detector.
Therefore it is necessary to confirm that we need another
 measurement of neutralino by direct or indirect experiment and also 
the mass and relic density should be consistent each other.

Moreover, the neutralino relic abundance, as determined at the LHC,
may come out convincingly outside the WMAP range.
 If $\abundchi$ comes out below WMAP range, several
solutions have been suggested which invoke non-standard cosmology,
e.g. quintessence-driven kination,
 while preserving the neutralino as the DM in the Universe.  
 However, if at
the same time direct and indirect DM searches bring null results, 
or even worse, the lightest super particle turns out to be charged particle
such as scalar tau,
this will provide a
strong indication against the neutralino nature of DM.
In fact, these inconsistencies can be perfectly explained
with axino or gravitino cold dark matter, which is dubbed as 
E-WIMPs~\cite{Choi:2005vq}.

Therefore this framework give us an opportunity to probe the features 
of the early Universe, since the relic density of E-WIMPs depends on
the reheating temperature $\treh$. 
Cosmic Lithium problems also can be solved with Gravitino 
DM~\cite{Jedamzik:2005dh}.

In this talk, based on ref.~\cite{Choi:2007rh}, we investigate the
determination of reheating temperature in the E-WIMP scenario using
the possible collider measurement of mass and relic density of NLSP, such as
neutralino or stau etc.

\section{E-WIMPs and Reheating temperature $\treh$}
  The spin-$1/2$ axino (the
fermionic superpartner of an axion) and the spin-$3/2$ gravitino (the
fermionic superpartner of a graviton) are both well-motivated E-WIMPs.
The former arises in SUSY extensions of models incorporating the
Peccei-Quinn solution to the strong CP problem. The latter is an
inherent ingredient of the particle spectrum of supergravity models.
 The characteristic
strength of their interactions with ordinary matter is strongly
suppressed by a large mass scale, the Peccei-Quinn scale $\fa\sim
10^{11}\gev$ in the case of axinos and the (reduced) Planck scale
$\mplanck\simeq 2.4\times 10^{18} \gev$ for gravitinos. The mass of
them are very model dependent and can vary  from keV up to TeV~\cite{ckn}
for axino and from eV to TeV for gravitino.
 In this work we want to remain as
model-independent as possible and will treat $\maxino$ and
$\mgravitino$ as free parameters.
 The possibility of
axinos as cold DM was pointed out in~\cite{ckr,ckkr}, 
while axinos as warm DM was considered in~\cite{rtw}. 
The heavy axino was studied in~\cite{Choi:2008zq}. 
The gravitino as a cosmological relic was
extensively studied in the literature. For more references refer 
to~\cite{Choi:2007rh}.
 
There are two generic ways to produce axinos or gravitinos.  
One proceeds via scatterings and decay processes of ordinary particles 
and sparticles in thermal bath. 
Its efficiency is proportional to their density in
the plasma which is a function of $\treh$ ({\em thermal production}). 
The other comes from  (out-of-equilibrium) decays of
the NLSPs, after their freeze-out, to E-WIMPs 
({\em non-thermal production}). 

The thermal production (TP) of axinos and gravitinos is a function of 
$\treh$~\cite{ckkr}.
For high $\treh$, both are almost proportional to $\treh$.
For axino~\cite{bs04}
\beqa{
\abundatp\simeq 5.5\, g_s^6 \ln \left(\frac{1.108}{g_s} \right)
\left(\frac{\maxino}{0.1\gev} \right)
\left(\frac{10^{11}\gev}{\fa}\right)^2 \left(\frac{\treh}{10^4 \gev} \right),
\label{eq:TP_axino_bs}
}
where $g_s$ is temperature-dependent strong coupling constant, which
in the above expression is evaluated at $\treh$. Note that,
$\yaxinotp\propto\treh/\fa^2$,
and for gravitino~\cite{bbb00}
\begin{equation}
\abundgtp\simeq 0.27 \left(\frac{\treh}{10^{10}\gev}\right)
\left(\frac{100\gev}{\mgravitino}\right) 
\left(\frac{\mgluino(\mu)}{1\tev}\right)^2,
\label{eq:abundgbbb}
\end{equation}
where $\mgluino(\mu)$ stands for the gluino
mass evaluated at a scale $\mu\simeq 1\tev$. 
 In the axino case, there is a sharp drop-off below
$\treh\sim1\tev$ due to Boltzmann suppression factor
$\exp{(-m/T)}$, with $m$ denoting here squark and gluino mass; 
at lower $\treh$ superpartner decay processes become dominant but are less
efficient~\cite{ckkr}.

For the non-thermal production (NTP), the relic abundance is 
simply given by
\beq
\abundlspntp = \frac{\mlsp}{\mnlsp} \abundnlsp.
\label{eq:ntp}
\eeq

The total abundance of the LSPs is the sum of both thermal and non-thermal 
production contributions
and it is natural to expect that the LSP makes up most of CDM in the
Universe, thus we can write 
\beqa{
\abundlsptp\left(\treh,\mlsp,\mgluino,\mnlsp,\ldots \right) +
\frac{\mlsp}{\mnlsp} \abundnlsp = \abundlsp = \abundcdm\simeq 0.1.
\label{eq:oh2relation}
}
Once the neutralino NLSP is discovered and its mass is determined at
the LHC with some precision, and so also $\abundnlsp=\abundchi$, then
eq.~(\ref{eq:oh2relation}) will provide a relation between $\treh$ and
$\mlsp$.

\section{Axino dark matter}
\begin{figure*}[t!]
\vspace*{-0.2in} 
  \begin{tabular}{c c}
    \includegraphics[width=0.45\textwidth]{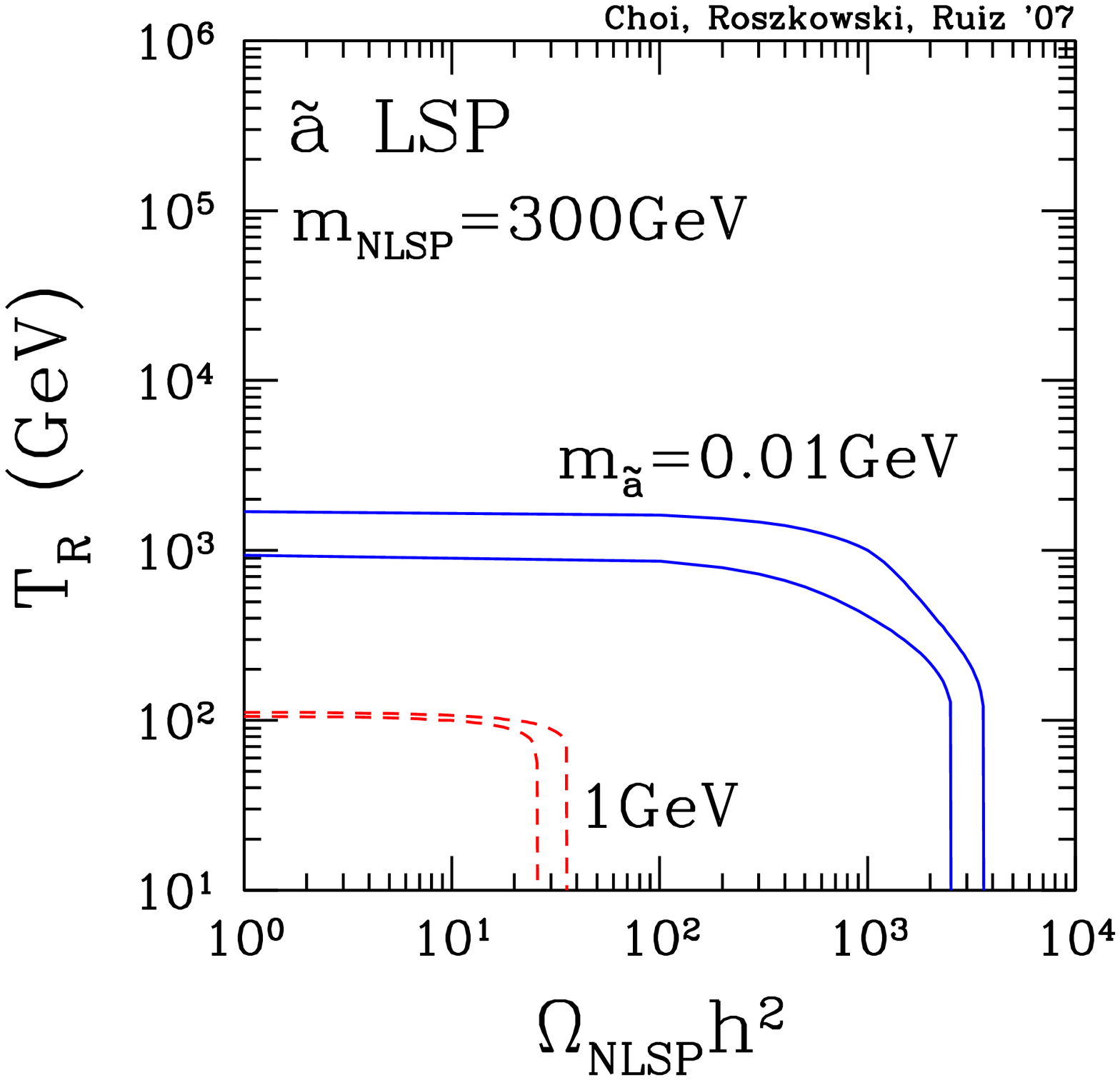}
&
    \includegraphics[width=0.45\textwidth]{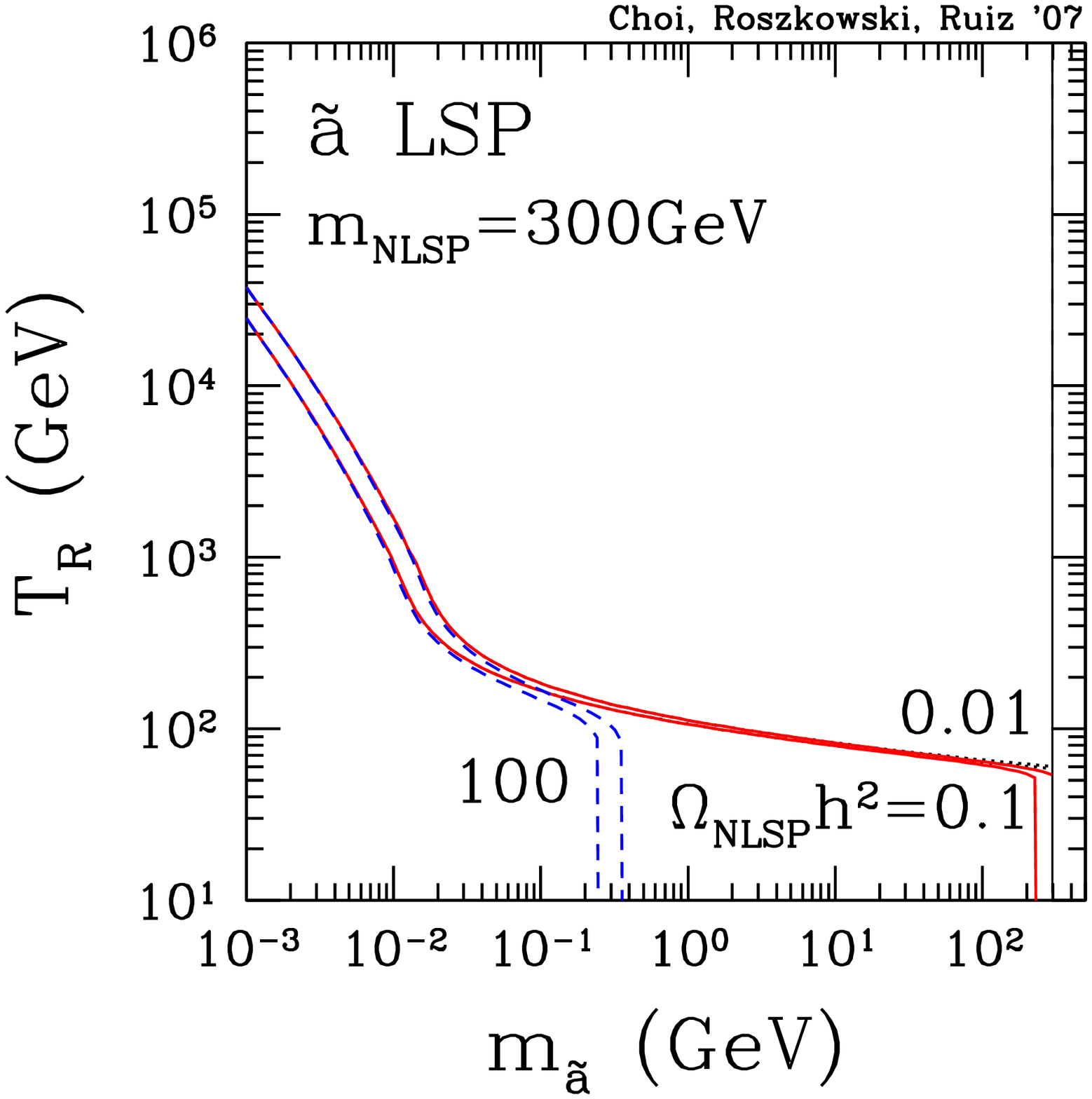}
    \end{tabular}
\caption{Left panel: $\treh$ vs. $\abundnlsp$ for $\mnlsp=300\gev$ and
for $\maxino=0.01\gev$ (solid blue) and $\maxino=1\gev$ (dashed
red). The bands correspond to the upper and lower limits of dark
matter density from WAMP.  Right panel: $\treh$
vs. $\maxino$ for $\abundnlsp=100$ (dashed blue), 0.1 (solid red) and
0.01 (dotted black). To the right of the solid vertical line the axino
is no longer the LSP. In both panels we set $\fa=10^{11}\gev$. }
\label{fig:axinotr}
\end{figure*}
\begin{figure}[!t]
  \begin{tabular}{c c}
    \includegraphics[width=0.45\textwidth]{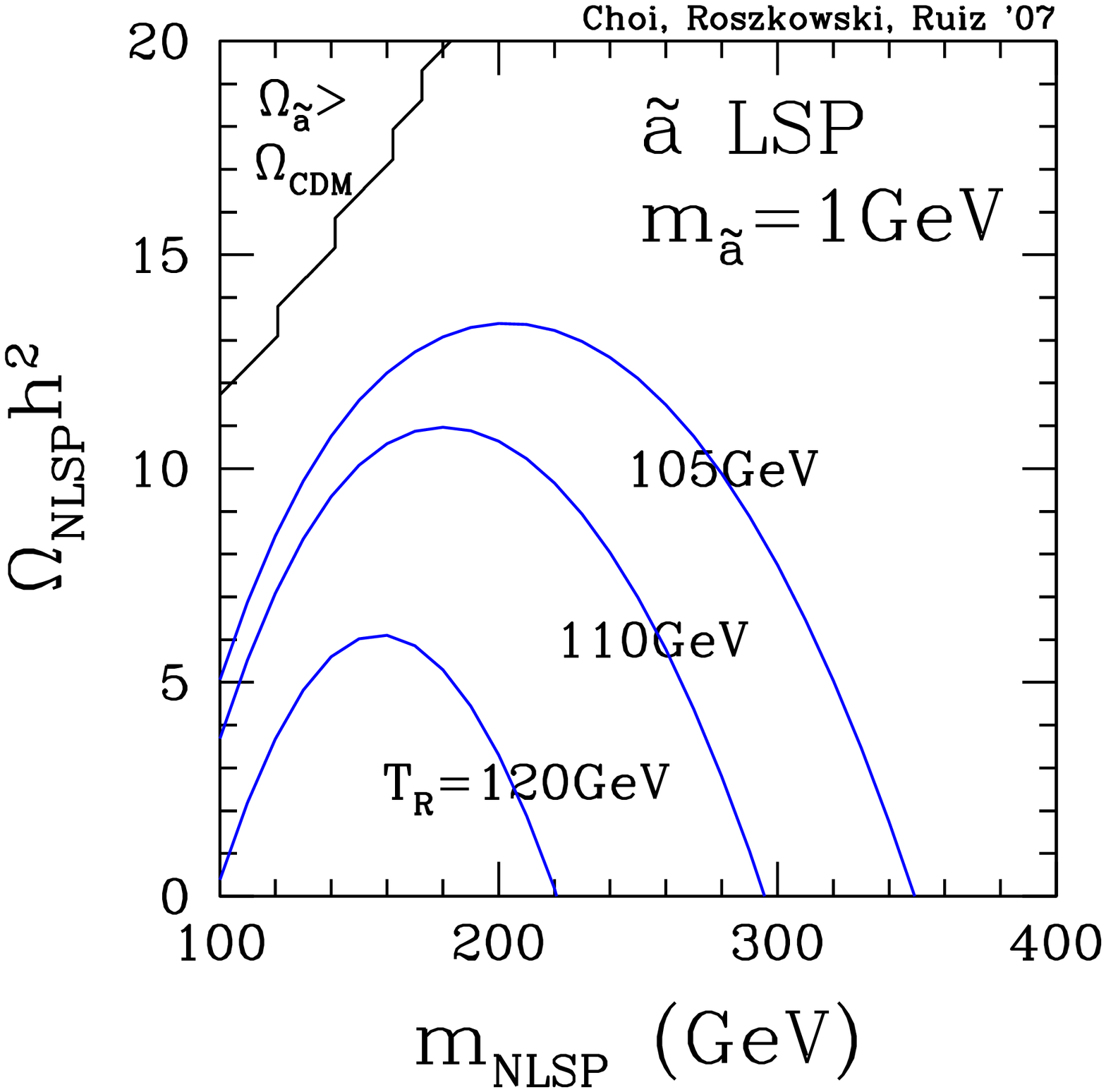}
&
    \includegraphics[width=0.45\textwidth]{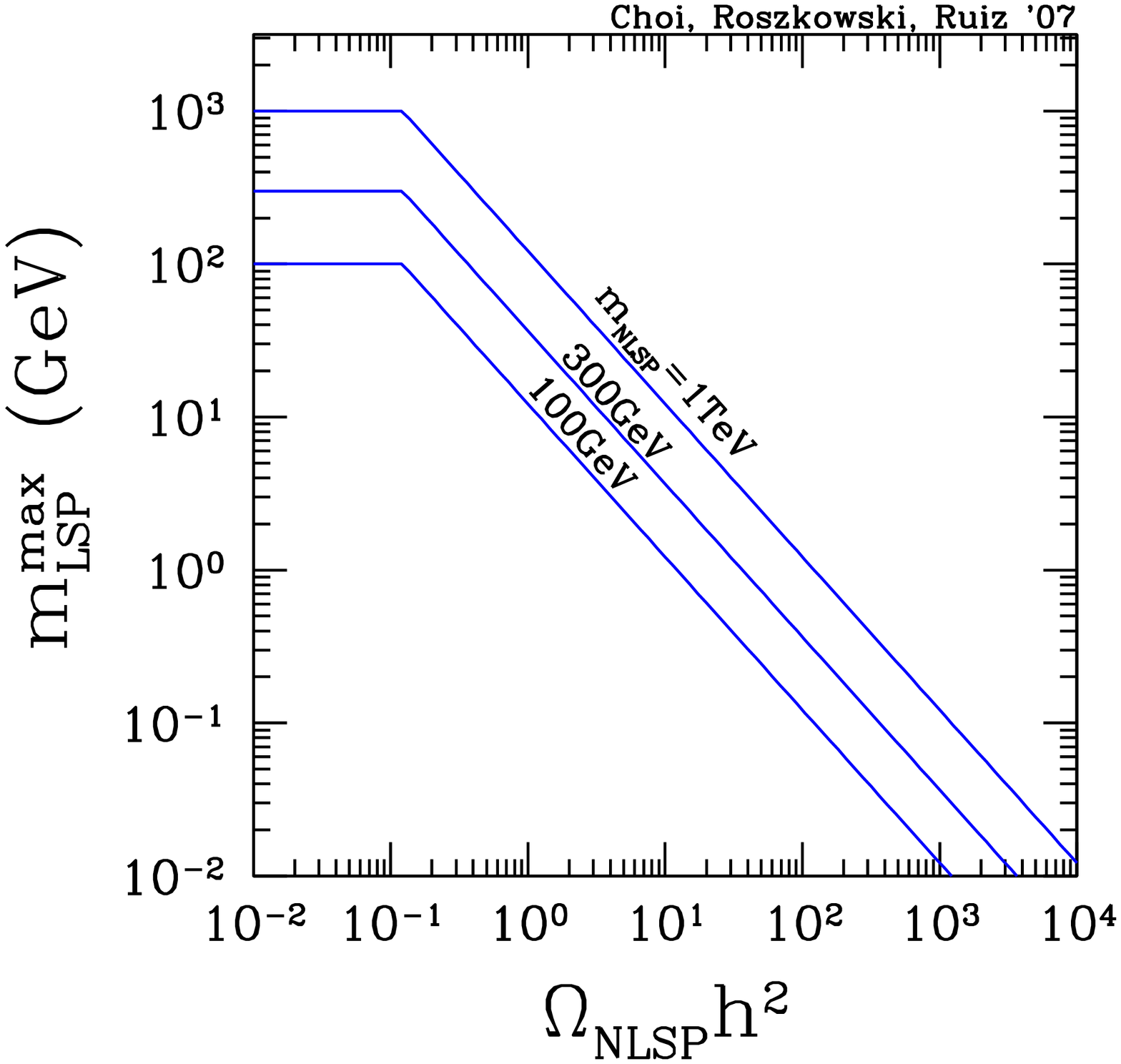}
    \end{tabular}
\caption{Left: Contours of the reheating temperature in the plane of
  $\mnlsp$ and $\abundnlsp$ such that $\abunda=\abundcdm=0.104$. The axino
  mass is assumed to be $1\gev$.
Right: Maximum values of $\mlsp$ as a function of $\abundnlsp$ for
  representative values of $\mnlsp$. Once both $\abundnlsp$ and
  $\mnlsp$ are determined from experiment, the upper bound on $\mlsp$
  can be derived. The plot applies both to the axino and to the
  gravitino LSP.}
\label{contour_axino1}
\end{figure}

First we consider axino as the LSP dark matter. 
Using (\ref{eq:oh2relation}), we can find the relations between the parameters,
$\treh$, $m_{LSP}$, $m_{NLSP}$, and $\abundnlsp$.
For fixed two parameters, we plot the contour on the space of the other 
two parameters in figures~\ref{fig:axinotr} and \ref{contour_axino1}.

From figure~\ref{fig:axinotr} (right panel), where $\maxino$ is small, 
TP dominates, $\abundatp \simeq \abundcdm$, hence we find
$\treh\propto \fa^2/\maxino$.
 This relation allows one to derive an {\em
upper bound} on $\treh$ if we use the fact that axinos have to be
heavy enough in order to constitute CDM. Assuming conservatively that
$\maxino\gsim100\kev$~\cite{ckkr}, we find $\trehmax<4.9\times10^5\gev$.

At larger $\maxino$ the NTP contribution becomes dominant and the
dependence on $\treh$ is lost, 
but in this regime the LSP mass becomes largest, this
allows one to derive an {\em upper bound} on $\maxino$. 
This is shown in fig.~\ref{contour_axino1} (right panel). 
Note that fig.~\ref{contour_axino1} (right panel) is actually applied to both
the axino and the gravitino LSP since it follows from
eq.~(\ref{eq:ntp}).

\section{Gravitino dark matter}
%
\begin{figure*}[t!]
\vspace*{-0.2in} 
  \begin{tabular}{c c}
    \includegraphics[width=0.45\textwidth]{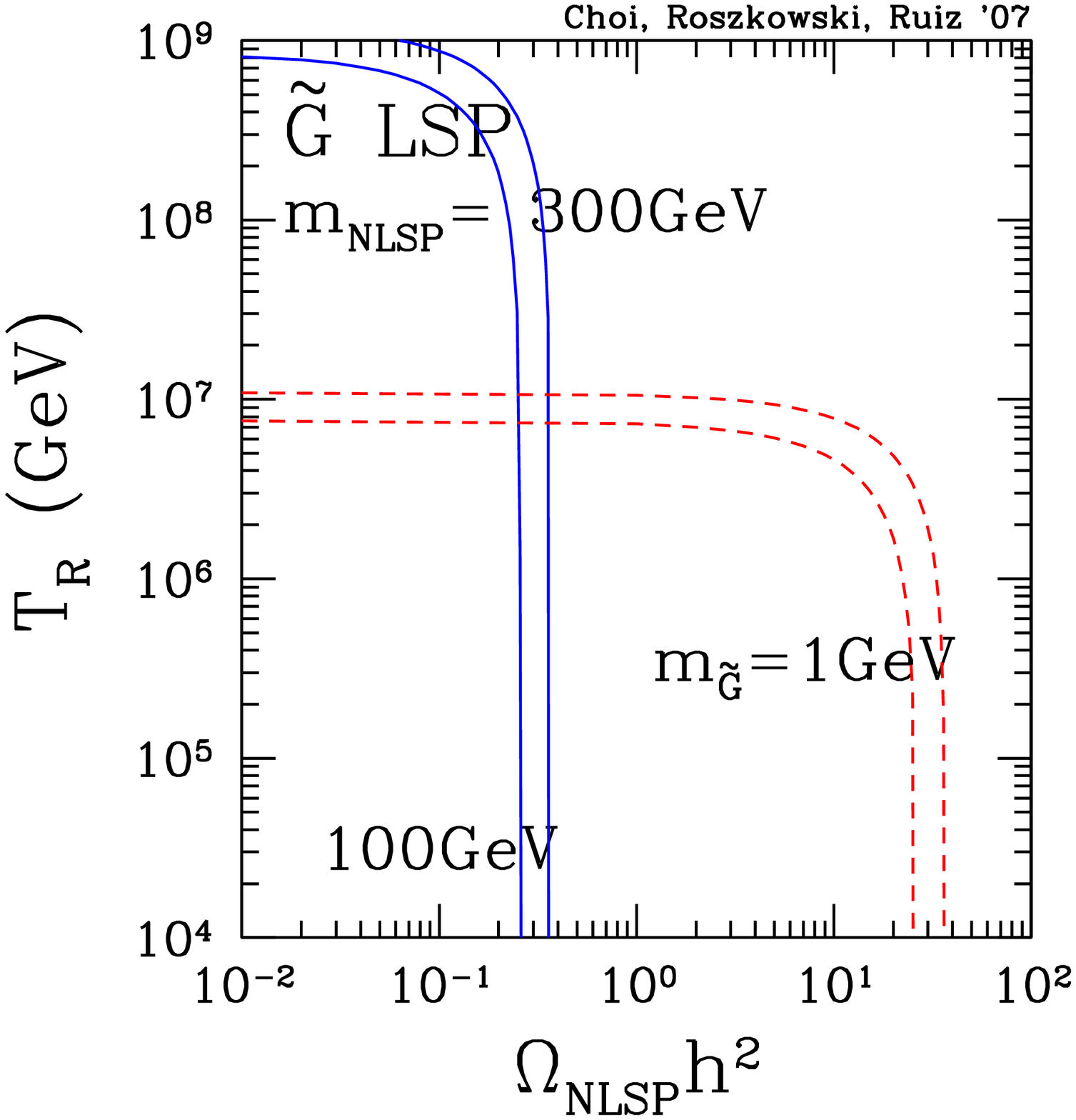}
&
    \includegraphics[width=0.45\textwidth]{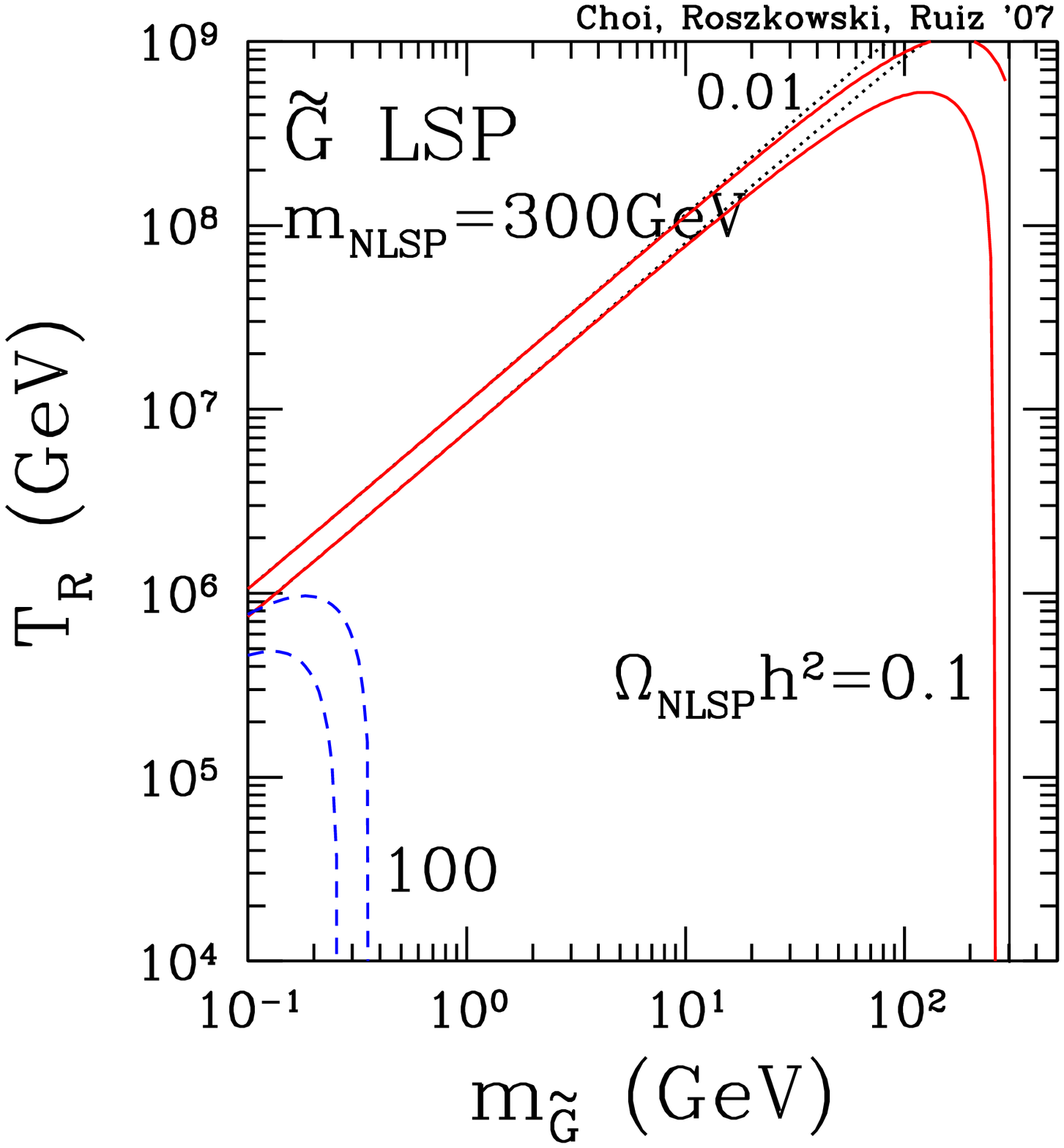}
    \end{tabular}
\caption{Left panel: $\treh$ vs. $\abundnlsp$ for $\mnlsp=300\gev$ and
for $\mgravitino=0.01\gev$ (solid blue) and $\mgravitino=1\gev$
(dashed red). The bands correspond to the upper and lower limits of
dark matter density from WMAP.  Right panel: $\treh$
vs. $\mgravitino$ for $\abundnlsp=100$ (dashed blue), 0.1 (solid red)
and 0.01 (dotted black).  To the right of the solid vertical line the
gravitino is no longer the LSP.}
\label{fig:gravitinotr}
\end{figure*}
\begin{figure*}[t!]
\vspace*{-0.2in} 
  \begin{tabular}{c c}
    \includegraphics[width=0.45\textwidth]{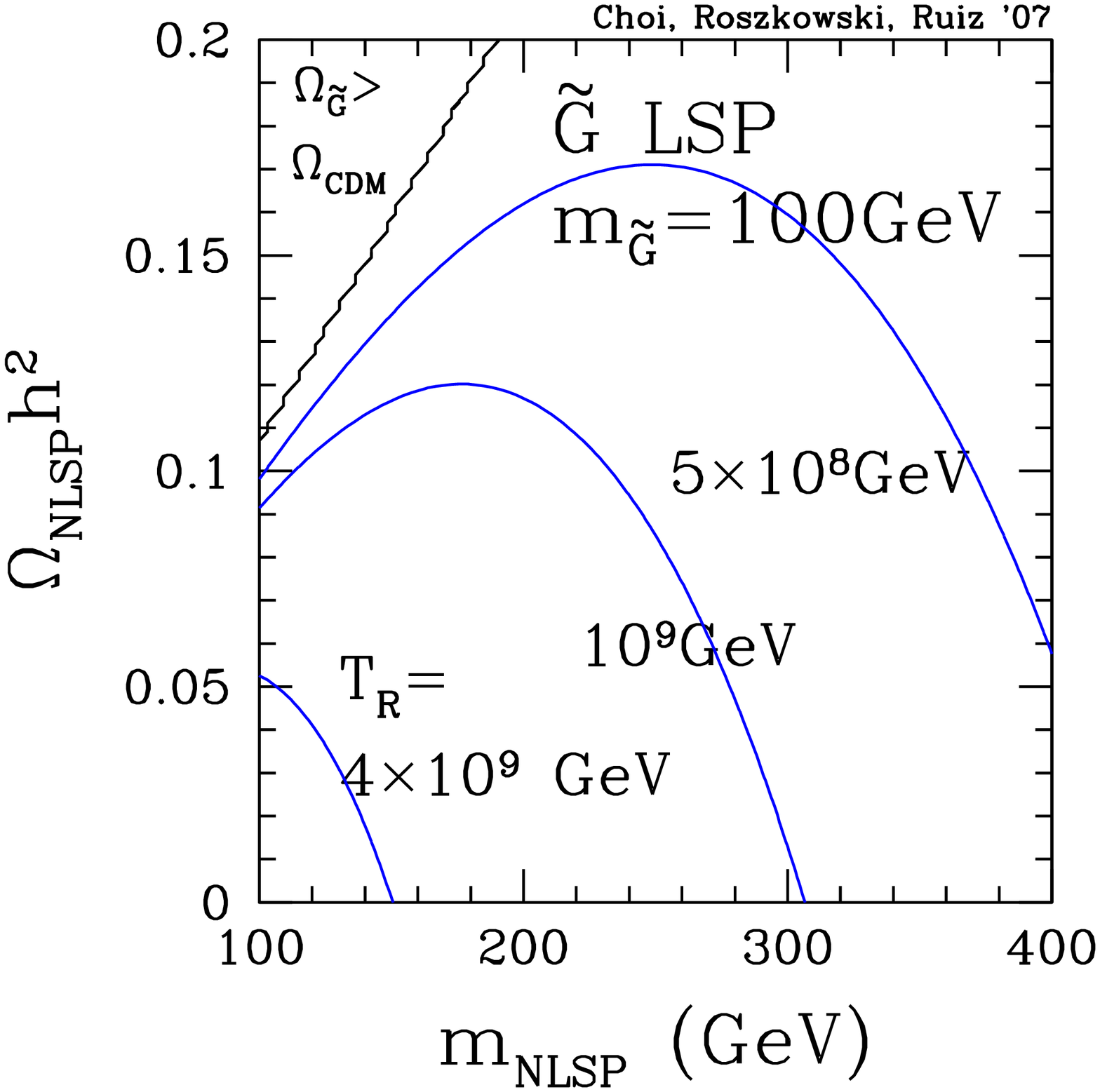}
&
    \includegraphics[width=0.45\textwidth]{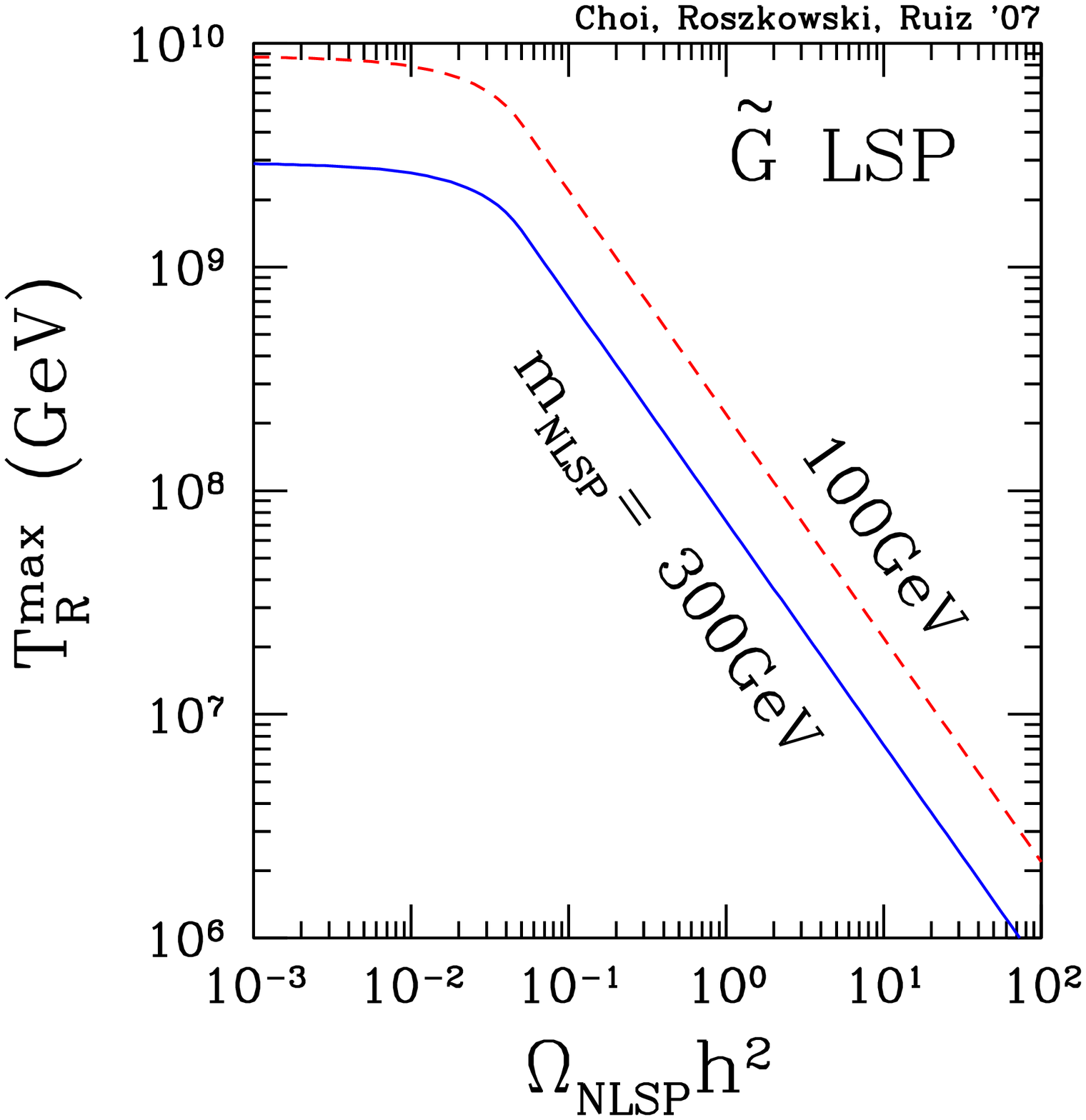}
    \end{tabular}
\caption{Left: Contours of the reheating temperature in the plane of
  $\mnlsp$ and $\abundnlsp$ such that $\abundg=\abundcdm=0.104$. The gravitino
  mass is assumed to be $100\gev$.
Right: Maximum reheating temperature $\trehmax$ vs. NLSP relic density 
$\abundnlsp$ with gravitino DM for NLSP mass $\mnlsp = 100 \gev$
  (dashed red) and $300 \gev$ (solid blue).}
\label{contour_gravitino1}
\end{figure*}

For gravitino LSP, the analogous plots are shown in 
figures~\ref{fig:gravitinotr} and~\ref{contour_gravitino1}.
For gravitino LSP, Big Bang Nucleosynthesis (BBN) gives strong constraint
since the lifetime of NLSP is around $1$ sec to $10^{12}$ sec due to the 
suppressed interaction. The neutralino NLSP is almost excluded with  
$\mgravitino\gsim 1 \gev$ due to BBN~\cite{fengetal,ccjrr}.

 While $\yaxinotp$ is independent
of the axino mass, in the  gravitino case $\ygravitinotp\propto
1/\mgravitino^2$. Thus $\abundatp\propto \maxino\treh$ while
$\abundgtp\propto \treh/\mgravitino$. In other words, if TP dominates,
$\abundgtp\simeq0.1$, we find
$\treh\propto\mgravitino$.
And then NTP dominates, $\treh$ drops down as shown 
in figure~\ref{fig:gravitinotr} (right panel). 
The turnover between the TP and NTP dominance
allows one to derive a conservative {\em
upper bound} $\trehmax$ which, unlike for the axino CDM, even without
knowing the gravitino mass. This is plotted in
fig.~\ref{contour_gravitino1} (right panel).

For stau NLSP case, in our numerical example in the right panel of
fig.~\ref{fig:gravitinotr}, with $\mstau=300\gev$, where we have also
taken $\mchi=477\gev$, the condition $\tau_{\stau}>10^3\sec$ implies
$\mgravitino\lsim2\gev$ and $\treh\lsim 9\times 10^6\gev$. Increasing
$\mstau$ to $1\tev$ and $\mchi$ to $1.5\tev$ leads to
$\mgravitino\lsim40\gev$ and $\treh\lsim 4\times 10^8\gev$. 
More detailed study  on $\treh$ bound with stau NLSP considering only 
thermal production and using the constraint from bound state
effects can be seen in~\cite{Steffen:2008bt}.

\section{Summary}
We studied the possible determination of reheating temperature
 with axino or gravitino LSP dark matter. 
We find that once we know the mass of NLSP and other parameters 
determining its relic
abundance from collider data we can give the reheating temperature
depending on the mass of LSP.
Even though the relic abundance of the NLSP is not measured precisely,
if the order of magnitude is much smaller or larger than WMAP range
then we can obtain conservative bound on the reheating temperature.

Note Added: Recently, and long after our work was published, a paper 
appeared~\cite{Steffen:2008bt} which, using a different set of variables 
(NLSP stau lifetime and mass, instead of our gravitino mass and NLSP mass) 
and neglecting non-thermal contribution to $\abund$, 
rederived several of our results. 
Specifically, with stau mass around $1 \tev$, an upper limit on the
reheating temperature of $\treh \lesssim 10^8 \gev$ was obtained, in
agreement with ours.

\begin{theacknowledgments}
K.-Y.C. is supported 
by the Ministerio de Educacion y Ciencia of Spain under
Proyecto Nacional FPA2006-05423 and by the Comunidad de Madrid under
Proyecto HEPHACOS, Ayudas de I+D S-0505/ESP-0346.
 L.R is partially supported by the EC
6th Framework Programmes MRTN-CT-2004-503369 and
MRTN-CT-2006-035505. R.RdA is supported by the program ``Juan de la
Cierva'' of the Ministerio de Educaci\'{o}n y Ciencia of Spain.

\end{theacknowledgments}


\end{document}